\documentclass[aps,prl,twocolumn,10pt,a4paper,superscriptaddress]{revtex4}
\pdfoutput=1

\usepackage{color}
\usepackage[latin1]{inputenc}
\usepackage{graphicx}
\usepackage{amsmath,amssymb}
\usepackage{booktabs}
\usepackage{hyperref}
\hypersetup{colorlinks=true,
						linkcolor=red,
						anchorcolor=blue,
						citecolor=blue,
						filecolor=blue,
						menucolor=blue,
						urlcolor=blue
}
\usepackage{color}

\renewcommand{\figurename}[1]{Figure }

\begin{document}
\title{Enhancement and sign change of magnetic correlations in a driven quantum many-body system}

\author{Frederik G\"org}
\affiliation{Institute for Quantum Electronics, ETH Zurich, 8093 Zurich, Switzerland}

\author{Michael Messer}
\affiliation{Institute for Quantum Electronics, ETH Zurich, 8093 Zurich, Switzerland}

\author{Kilian Sandholzer}
\affiliation{Institute for Quantum Electronics, ETH Zurich, 8093 Zurich, Switzerland}

\author{Gregor Jotzu}
\affiliation{Institute for Quantum Electronics, ETH Zurich, 8093 Zurich, Switzerland}
\affiliation{Max Planck Institute for the Structure and Dynamics of Matter, 22761 Hamburg, Germany}

\author{R\'emi Desbuquois}
\affiliation{Institute for Quantum Electronics, ETH Zurich, 8093 Zurich, Switzerland}

\author{Tilman Esslinger}
\affiliation{Institute for Quantum Electronics, ETH Zurich, 8093 Zurich, Switzerland}

\maketitle

\textbf{
Periodic driving can be used to coherently control the properties of a many-body state and to realize new phases which are not accessible in static systems. 
For example, exposing materials to intense laser pulses enables to provoke metal-insulator transitions, control the magnetic order and induce transient superconducting behavior well above the static transition temperature\;\cite{Kirilyuk2010,Nicoletti2016,Rini2007,Mariager2012,Li2013,Mitrano2016}. 
However, pinning down the responsible mechanisms is often difficult, since the response to irradiation is governed by complex many-body dynamics. 
In contrast to static systems, where extensive calculations have been performed to explain phenomena such as high-temperature superconductivity\;\cite{Dagotto1994}, theoretical analyses of driven many-body Hamiltonians are more demanding and new theoretical approaches have been inspired by the recent observations\;\cite{Mentink2015,Coulthard2016,Kitamura2016}. 
Here, we perform an experimental quantum simulation in a periodically modulated hexagonal lattice and show that anti-ferromagnetic correlations in a fermionic many-body system can be reduced or enhanced or even switched to ferromagnetic correlations. 
We first demonstrate that in the high frequency regime, the description of the many-body system by an effective Floquet-Hamiltonian with a renormalized tunneling energy remains valid, by comparing the results to measurements in an equivalent static lattice. 
For near-resonant driving, the enhancement and sign reversal of correlations is explained by a microscopic model, in which the particle tunneling and magnetic exchange energies can be controlled independently. 
In combination with the observed sufficiently long lifetime of correlations, Floquet engineering thus constitutes an alternative route to experimentally investigate unconventional pairing in strongly correlated systems\;\cite{Dagotto1994,Coulthard2016,Kitamura2016}. 
}

The increasing demand for high speed control of current magnetic memory devices in the terahertz regime has led to efforts to optically control the magnetic properties of materials, for example to switch from anti-ferromagnetic to ferromagnetic ordering\;\cite{Mariager2012,Li2013}. 
To engineer suitable materials for future applications, it is desirable to gain a better understanding of the underlying microscopic processes. 
In this context, cold atom experiments provide an ideal platform to investigate driven many-body systems due to the slow timescales and the prospect of quantitative comparisons to theoretical predictions. 
So far, periodic modulation has been employed in such setups to engineer effective Hamiltonians\;\cite{Goldman2014a,Bukov2015b}, which enabled to renormalize Hubbard parameters and study classical magnetism in the high-frequency regime as well as to realize new features like topological or spin-dependent bandstructures\;\cite{Struck2011,Jotzu2015,Eckardt2017}. 
By driving interacting systems\;\cite{Zenesini2009,Parker2013}, both charge and spin degrees of freedom can be influenced by individually addressing density-dependent processes\;\cite{Ma2011,Meinert2016,Desbuquois2017}. 
Until now, the measurement of magnetic correlations in driven optical lattices has remained an open challenge. 
An experimental difficulty lies in the heating associated with the periodic modulation of a many-body system which may destroy correlations, in particular in the near-resonant regime\;\cite{Jotzu2015,Kuwahara2016,Abanin2017}. 

\begin{figure}
	\includegraphics[width=1\columnwidth]{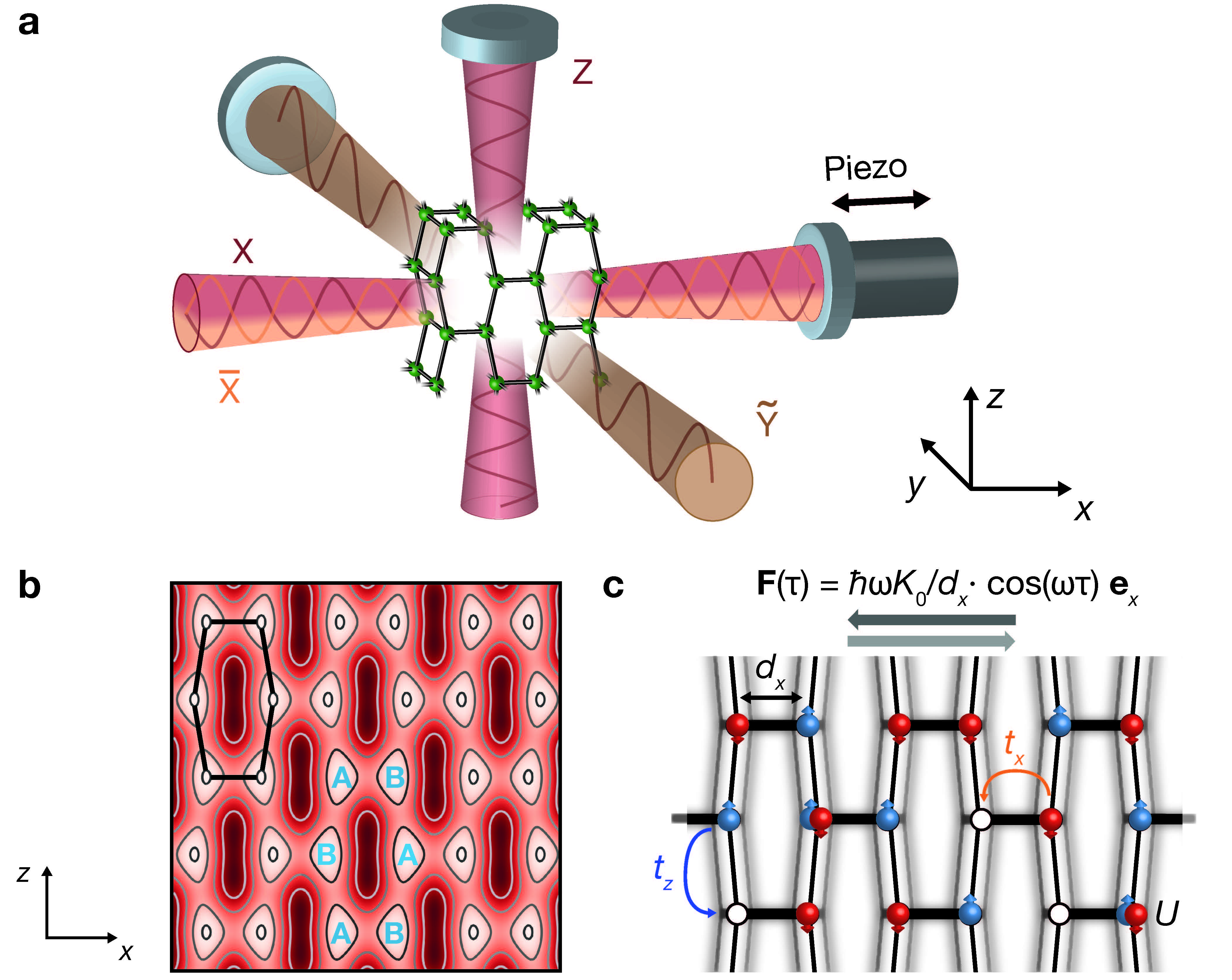}
	\caption{
	\textbf{Experimental setup.} 
	\textbf{a,} 
	Optical lattice setup to create the three-dimensional geometry. The beams $\text{X}$ and $\text{Z}$ are interfering, while $\overline{\text{X}}$ and $\widetilde{\text{Y}}$ are frequency-detuned. A piezoelectric actuator sinusoidally modulates the position of the retroreflecting mirror in $x$-direction.
	\textbf{b,}
	Lattice potential in the $x$-$z$-plane consisting of $\mathcal{A}$ and $\mathcal{B}$ sublattices with superimposed hexagonal unit cell.
	\textbf{c,}
	Tight-binding representation of the lattice potential in the $x$-$z$-plane. The system is described by a driven Fermi-Hubbard model, with anisotropic tunneling energies $t_x>t_z$ due to a shorter length $d_x$ of the horizontal bonds. Atoms in different spin states interact via an onsite interaction $U$. In a co-moving frame, the modulation of the lattice position corresponds to a linear force $\mathbf{F}(\tau)$ in $x$-direction with an amplitude $\hbar\omega K_0/d_x$, which primarily influences the horizontal bonds (Methods). 
	}
	\label{fig:1}
\end{figure} 

We perform our experiments with a degenerate Fermi gas consisting of $3.0(2)\times 10^4$ (10\% systematic error) ultracold $^{40}\text{K}$ atoms prepared in a balanced mixture of two internal states, denoted as $\uparrow$ and $\downarrow$ (see Methods). 
The atoms are loaded into a tunable geometry optical superlattice with anisotropic tunneling rates, where the horizontal links in $x$-direction $t_x$ are stronger than in the $y$- and $z$-directions $t_{y,z}$ (see Fig.\;1c). 
In the $x$-$z$ plane, the lattice consists of hexagonal layers, which are stacked in the $y$-direction. 
We modulate the lattice position in $x$-direction periodically in time with a displacement amplitude $A$ at a frequency $\omega/2\pi$, which is achieved by moving the retroreflecting mirror of the optical lattice with a piezoelectric actuator (see Fig.\;1a). 

Our system is well described by the driven Fermi-Hubbard model
\begin{equation}
\hat{H}(\tau)=-\sum_{\substack{\left\langle \mathbf{i},\mathbf{j}\right\rangle \\ \sigma}} t_{\mathbf{ij}} c^{\dagger}_{\mathbf{i}\sigma}c_{\mathbf{j}\sigma}+U \sum_\mathbf{i} \hat{n}_{\mathbf{i}\uparrow}\hat{n}_{\mathbf{i}\downarrow}+\sum_{\mathbf{i},\sigma} \left(f_{\mathbf{i}}(\tau)+V_\mathbf{i}\right) \hat{n}_{\mathbf{i}\sigma}
\label{drivenFHM}
\end{equation}
where $c^{\dagger}_{\mathbf{i}\sigma}$ and $\hat{n}_{\mathbf{i}\sigma}$ are the fermionic creation and number operators at site $\mathbf{i}=(i_x,i_y,i_z)$ in spin-state $\sigma=\uparrow,\downarrow$, respectively. 
Here, $t_{\mathbf{ij}}$ denotes the tunneling rate between nearest neighbors $\left\langle \mathbf{i},\mathbf{j}\right\rangle$, $U$ the repulsive onsite interaction and $V_\mathbf{i}$ an overall harmonic trapping potential. 
The time-dependent force can be expressed as $f_{\mathbf{i}}(\tau)=m A\omega^2 x_{\mathbf{i}}\cos(\omega\tau)$, where $m$ is the mass of the atoms and $x_{\mathbf{i}}=\left\langle \hat{x}\right\rangle_{\mathbf{i}}$ the $x$-position of the Wannier function on site $\mathbf{i}$. 
Therefore, the driving can be used to primarily address the bonds in the $x$-direction (Methods). 
To characterize the many-body state in the lattice, we measure the fraction of atoms on doubly occupied sites 
\begin{equation}
\mathcal{D}=\frac{2}{N} \sum_{\mathbf{i}\in \mathcal{A},\mathcal{B}} \left\langle \hat{n}_{\mathbf{i}\uparrow}\hat{n}_{\mathbf{i}\downarrow}\right\rangle \notag
\end{equation}
as well as the nearest neighbor spin-spin correlator 
\begin{equation}
\mathcal{C}=-\frac{1}{N} \sum_{\mathbf{i}\in \mathcal{A}}(\langle \hat{S}_{\mathbf{i}}^x \hat{S}_{\mathbf{i}+\mathbf{e}_x}^x\rangle+\langle \hat{S}_{\mathbf{i}}^y \hat{S}_{\mathbf{i}+\mathbf{e}_x}^y\rangle) \notag
\end{equation}
on the horizontal links along the $x$-direction ($N$ is the total atom number and $\mathbf{e}_x$ the unit vector in $x$-direction). 
The observables are averaged spatially over the inhomogeneous density distribution in the harmonic trap with a geometric mean trapping frequency of $\bar{\omega}_{\text{trap}}/2\pi=84(2)\;\text{Hz}$ as well as over one oscillation cycle of the periodic modulation as indicated by $\left\langle ...\right\rangle$ (see Methods). 

In a first experiment, we investigate the regime where the driving frequency is much higher than all microscopic energy scales of the system, i.e. the tunneling and interaction energies ($\hbar\omega \gg t,U$). 
In the non-interacting case, the modulation renormalizes the horizontal tunneling rate by a zeroth order Bessel function and the system can be described by an effective tunneling energy
\begin{equation}
t_x^{\text{eff}}(K_0)=t_x \mathcal{J}_0(K_0),
\label{TunOffResDrive}
\end{equation}
where $K_0=m\,A\,\omega\,d_x/\hbar$  is the normalized drive amplitude, with $d_x$ the length of the horizontal bonds (see Fig.\;1c)\;\cite{Eckardt2017}. 
However, it is not a priori clear if this simple description remains accurate in the many-body context\;\cite{Bukov2015b}. 
To verify this, we compare our measurements in the driven system to results obtained using an experimental quantum simulation in a static lattice with a variable tunneling rate $t_x$. 
The reliability of our experiment as a quantum simulator for the magnetic properties of the Hubbard model has previously been benchmarked through quantitative comparisons with state-of-the-art numerical calculations\;\cite{Sciolla2013,Imriska2014}. 
To enter the driven regime in the experiment, we linearly ramp up the lattice modulation amplitude to a final value $K_0$ within $2\;\text{ms}$, at a frequency of $\omega/2\pi=6\;\text{kHz}$. 
Afterwards, we allow for an additional equilibration time of $5\;\text{ms}$ before the measurement, during which we maintain a fixed modulation amplitude. 

The resulting double occupancies and spin correlations agree well for the driven and static cases, as shown in Fig.\;2. 
This supports the validity of the description of the many-body system by an effective Hamiltonian with a tunneling rate $t_x^{\text{eff}}(K_0)$. 
For lower tunneling energies, the double occupancy decreases due to the reduction of the bandwidth $W$. 
Therefore, for increasing driving amplitude, the system is entering the Mott regime\;\cite{Zenesini2009}. 
The modulation not only changes the bandwidth, but also the anisotropy of the lattice, since the ratio $t_x^{\text{eff}}(K_0)/t_{y,z}$ decreases for increasing driving amplitude. 
This effect manifests itself in the spin correlator on the horizontal link, which decreases for a weaker anisotropy of the underlying lattice, as observed in previous measurements\;\cite{Greif2015}. 
When driving for longer times, we find that the lifetime of correlations is reduced to $14(5)\;\text{ms}$ at $K_0=1.26(4)$ compared to $92(16)\;\text{ms}$ in the static case. 
Nevertheless, this allows to observe comparable levels of correlations in the driven and static cases on experimental timescales. 

\begin{figure}
	\includegraphics[width=1\columnwidth]{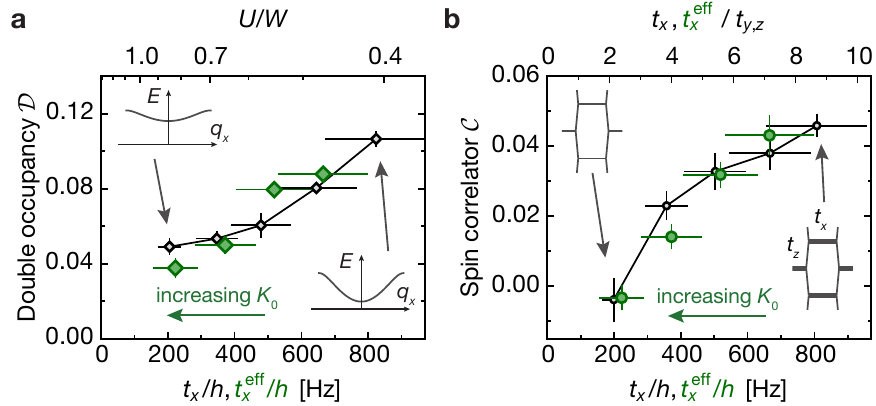}
	\caption{
	\textbf{Description of the driven system by an effective Hamiltonian in the high-frequency regime.} 
	\textbf{a,} 
	Double occupancy $\cal{D}$  as a function of effective horizontal tunneling energy $t_x^{\text{eff}}(K_0)=t_x \mathcal{J}_0(K_0)$ for a driven system (green) and results obtained through an experimental quantum simulation in a static configuration with horizontal tunneling $t_x$ (black). The inset shows a cut through the non-interacting bandstructure as a function of the quasi-momentum in $x$-direction $q_x$. The reduction of the bandwidth $W$ leads to a lower double occupancy, indicating the crossover to a Mott-insulating state.
	\textbf{b,} Spin correlations $\cal{C}$ as a function of the (effective) horizontal tunneling energy for the driven case (green) and an equivalent static configuration (black). The renormalization of the tunneling energy leads to a reduction in lattice anisotropy $t_x^{\text{eff}}/t_{y,z}$ (see inset), which reduces the magnetic correlations on the horizontal link. 
	The transverse tunneling energies are $t_y/h=125(8)\;\text{Hz}$ and $t_z/h=78(8)\;\text{Hz}$ and the interaction is set to $U/h=0.93(2)\;\text{kHz}$. 
	Horizontal error bars reflect the uncertainty in the lattice depth, data points and vertical error bars in \textbf{a} (\textbf{b}) denote the mean and standard error of 4 (10) individual measurements at different times withing one driving period (see Methods). 
	}
	\label{fig:2}
\end{figure}

While an off-resonant modulation scheme typically leads to a renormalization of pre-existing parameters, novel physics which is not accessible in static systems arises for a near-resonant drive. 
For example, extended terms such as density-dependent tunneling energies can be engineered, which are not present in the single band Hubbard model\;\cite{Ma2011,Meinert2016,Desbuquois2017}. 
To investigate this regime, we set a large onsite interaction close to the driving frequency $U\approx l\hbar\omega$ ($l\in \mathbb{Z}$) and ramp up the modulation at a frequency of either $3\;\text{kHz}$ or $6\;\text{kHz}$ within $3.3\;\text{ms}$ or $2\;\text{ms}$, respectively. 
We observe that the effective states in the driven Hamiltonian contain a higher fraction of double occupancies if $U\approx l\hbar\omega$, see Fig.\;3a. 

\begin{figure}
	\includegraphics[width=1\columnwidth]{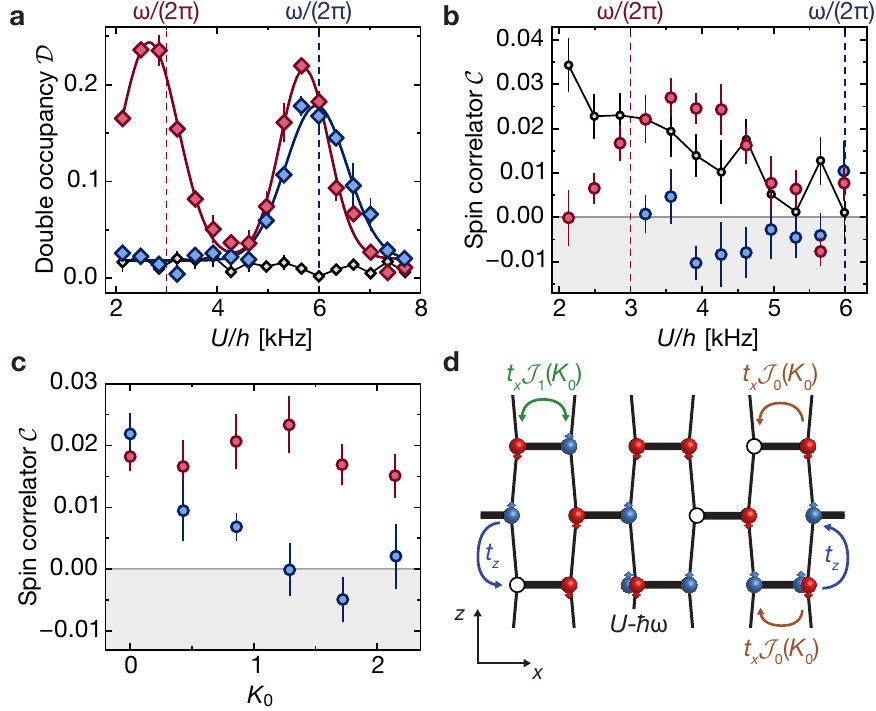}
	\caption{
	\textbf{Enhancement and sign reversal of magnetic correlations by near-resonant driving.}
	\textbf{a,} 
	Double occupancy as a function of onsite interaction $U$ for the static case (black) and driving frequencies of $\omega/2\pi=3\;\text{kHz}$ (red) and $6\;\text{kHz}$ (blue) with a modulation amplitude of $K_0=1.30(3)$. Around the resonances, the effective states in the driven Hamiltonian contain a higher number of double occupancies. Solid lines are (double) Gaussian fits to the data. 
	\textbf{b,}
	Spin correlations on the horizontal link as a function of $U$ for the same parameters as in \textbf{a}. For $U>\hbar\omega$ (red), anti-ferromagnetic correlations are enhanced compared to the static case (black) for a broad range of interactions. When $U<\hbar\omega$ (blue), the correlator changes sign and the system develops ferromagnetic correlations.
	\textbf{c,}
	Spin correlations as a function of driving amplitude $K_0$ for $\omega/2\pi=3\;\text{kHz}$, $U/h=3.8(1)\;\text{kHz}$ (red) and $\omega/2\pi=6\;\text{kHz}$, $U/h=4.4(1)\;\text{kHz}$ (blue). For $U>\hbar\omega$, anti-ferromagnetic correlations increase around $K_0\approx 1.3$. For $\hbar\omega >U$, correlations become ferromagnetic beyond a critical modulation amplitude.
		The tunneling rates are set to $(t_x,t_y,t_z)/h=(570(110),125(8),85(8))\;\text{Hz}$. 
		Data points and error bars in \textbf{a} (\textbf{b},\textbf{c}) denote the mean and standard error of 4 (10) individual measurements at different times withing one driving period (see Methods). 
	\textbf{d,}
	In the near-resonant case $U\approx \hbar\omega$, the driven system can be described by an effective Hamiltonian in which tunneling processes that do not change the number of double occupancies are renormalized by $\mathcal{J}_0(K_0)$. In contrast, the creation of doublon-holon pairs is resonantly enhanced and is determined by the first order Bessel function $\mathcal{J}_1(K_0)$. The effective interaction of the system becomes $U-\hbar\omega$.
	}
	\label{fig:3}
\end{figure} 

Strikingly, we find that the magnetic correlations on the horizontal links depend both on the sign and magnitude of the modulation detuning $\delta=\hbar\omega-U$, see Fig.\;3b. 
For a red-detuned drive ($\delta<0$), correlations are increased compared to the static case if $\left|\delta\right|$ is on the order of a few tunneling energies $t_x$. 
In contrast, when choosing $\delta>0$ the sign of the spin-spin correlator inverts, i.e. the system exhibits ferromagnetic correlations on neighboring sites in the $x$-direction. 
If we set a fixed interaction strength and vary the amplitude of the modulation, we find that for $\delta<0$ correlations increase for values around $K_0\approx 1.3$ before they eventually decrease again (see Fig.\;3c). 
For $\delta>0$ a critical value of the driving strength is required for the system to develop ferromagnetic correlations. 
We also study the time-dependence of the magnetic properties by varying the modulation time after the ramp up of the drive. 
We find that it takes a few milliseconds until correlations increase or change sign, respectively, before they approach zero when driving for long times due to heating of the cloud (see extended data Fig.\;ED1). 
The lifetime of magnetic correlations as extracted from an exponential fit to the long-time behavior changes from $82(34)\;\text{ms}$ in the static case to $12(4)\;\text{ms}$ at $K_0=1.30(3)$. 
In addition, we also observe the fast dynamics within one period of the drive (the so called micromotion) in our measurement regime (see extended data Fig.\;ED2). 
Finally, we investigate the adiabaticity of the preparation protocol by reverting the driving ramp and find that correlations return only partially to their static values (see extended data Fig.\;ED3). 

In order to obtain a microscopic understanding of the observed phenomena, we perform a Floquet analysis on the time-periodic Hamiltonian (1) in the near-resonant driving regime with $t\ll U\approx l\hbar\omega$. 
For that, we go to a rotating frame with respect to the operator 
\begin{equation}
\hat{R}(\tau)=\exp[i\sum_{\mathbf{j}}(l\omega\tau\hat{n}_{\mathbf{j}\uparrow}\hat{n}_{\mathbf{j}\downarrow}+\sum_{\sigma} F_{\mathbf{j}}(\tau)\hat{n}_{\mathbf{j}\sigma})], \notag
\end{equation}
where 
\begin{equation}
F_{\mathbf{j}}(\tau)=1/\hbar\int_0^{\tau} f_{\mathbf{j}}(\tau')d\tau'. \notag
\end{equation}
In this frame, the tunneling on the horizontal bonds is to lowest order in $1/\omega$ described by the effective Hamiltonian
\begin{equation}
\hat{H}^{\text{eff}}_{t_x}=-t_x \sum_{\substack{\mathbf{i}\in A,\sigma \\ \mathbf{j}=\mathbf{i}+\mathbf{e}_x}} \left[ \mathcal{J}_0(K_0) \hat{a}_{\mathbf{ij}\bar{\sigma}} + \mathcal{J}_l(K_0) \hat{b}^l_{\mathbf{ij}\bar{\sigma}} \right]c^{\dagger}_{\mathbf{i}\sigma}c_{\mathbf{j}\sigma}+\text{h.c.}
\label{TunResDrive}
\end{equation}
where $\bar{\uparrow}=\;\downarrow$ and vice versa\;\cite{Bermudez2015,Itin2015,Bukov2016}. Here, the effective tunneling energy is density-dependent: Hopping processes which do not change the number of double occupancies as described by the operator $\hat{a}_{\mathbf{ij}\sigma}=(1-\hat{n}_{\mathbf{i}\sigma})(1-\hat{n}_{\mathbf{j}\sigma})+\hat{n}_{\mathbf{i}\sigma}\hat{n}_{\mathbf{j}\sigma}$ are renormalized by $\mathcal{J}_0(K_0)$. 
In contrast, the creation or annihilation of doublon-holon pairs corresponding to $\hat{b}^l_{\mathbf{ij}\sigma}=(-1)^l (1-\hat{n}_{\mathbf{i}\sigma})\hat{n}_{\mathbf{j}\sigma}+\hat{n}_{\mathbf{i}\sigma}(1-\hat{n}_{\mathbf{j}\sigma})$ become resonantly restored with an amplitude $t_x \mathcal{J}_l(K_0)$ (see Fig.\;3d). 
In addition, the effective interaction $U^{\text{eff}}=U-l\hbar\omega=-\delta_l$ is given by the detuning from the $l$-photon resonance $\delta_l$. 
In this picture, one can understand the creation of double occupancies for small $\delta_l$ shown in Fig.\;3a as the system becoming effectively more weakly interacting. 

The magnetic properties of the many-body state are significantly altered in the effective Hamiltonian (3), since microscopically the superexchange process leading to spin-spin interactions involves two virtual hopping processes determined by $\mathcal{J}_l(K_0)$, in which a double occupancy at energy $U^{\text{eff}}$ is created and annihilated. 
Therefore, the exchange energy $J_{\text{ex}}$, which is the energy splitting between a spin singlet and triplet state on the horizontal bonds, will depend both on the modulation amplitude $K_0$ and the detuning $\delta$. 
It can even change sign for $\delta>0$, since in this case the effective interaction becomes attractive\;\cite{Trotzky2008,Mentink2015,Coulthard2016,Kitamura2016} (see extended data Fig.\;ED4 and Methods). 
We directly measure $J_{\text{ex}}$ between neighboring sites in the experiment using our tunable geometry optical lattice. 
For that, we disconnect individual pairs of sites in the $x$-direction from each other by raising the potential barrier in the $y$- and $z$-directions, such that the coupling $t_{y,z}/h<2\;\text{Hz}$ becomes negligible, and measure the exchange energy in a Ramsey-type sequence (see Fig.\;4a)\;\cite{Trotzky2008,Chen2011}. 

\begin{figure}
	\includegraphics[width=1\columnwidth]{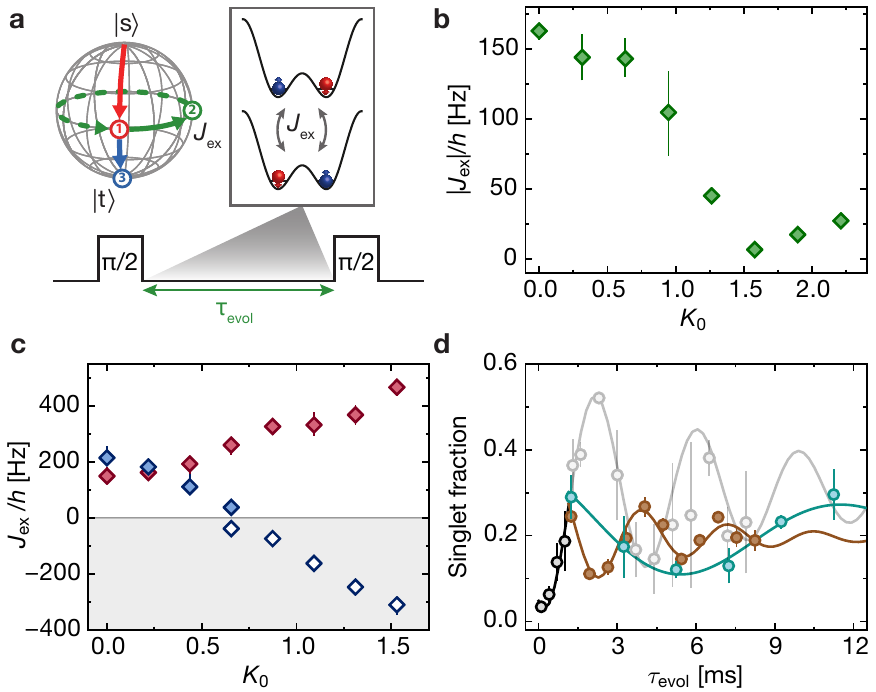}
	\caption{
	\textbf{Magnetic exchange energy for off- and near-resonant driving.} 
	\textbf{a,} 
	The exchange $J_\text{ex}$ is measured by preparing local singlet states $\left|\text{s}\right\rangle$ on isolated double wells. In a Ramsey-type sequence, a superposition between the singlet and triplet state $\left|\text{t}\right\rangle$ is created by performing a $\pi/2$-pulse with a magnetic field gradient. The exchange oscillation is triggered by suddenly lowering the barrier in the double well. After a variable evolution time $\tau_{\text{evol}}$, a second $\pi/2$-pulse is applied and the final singlet fraction is measured, which oscillates at a frequency $\left|J_{\text{ex}}\right|$. 
	\textbf{b,} 
	Magnetic exchange in the off-resonant driving regime for $\omega/2\pi=8\;\text{kHz}$, $t_x/h=350(50)\;\text{Hz}$ and $U/h=2.1(1)\;\text{kHz}$ as a function of driving amplitude. $J_\text{ex}$ decreases with $K_0$ as expected for a renormalized tunneling rate $t_x^{\text{eff}}$.
	\textbf{c,} 
	Exchange energy for near-resonant modulation with $\omega/2\pi=8\;\text{kHz}$, $t_x/h=640(90)\;\text{Hz}$ and $U/h=9.1(1)\;\text{kHz}$ (red) or $U/h= 6.5(1)\;\text{kHz}$ (blue), respectively, as a function of $K_0$. Red detuned driving ($U>\hbar\omega$) enhances the magnetic exchange for increasing driving amplitude. For $U<\hbar\omega$, $J_{\text{ex}}$ vanishes at a critical value $K_0\approx 0.7$ and becomes negative for stronger driving (open symbols). The sign of the exchange is measured as shown in \textbf{d}. For $K_0\approx 0.7$, the oscillation is too slow to determine the sign of $J_{\text{ex}}$. Mean values are derived from a sinusoidal fit to the data, errors denote the standard deviation obtained from a resampling method (see Methods). 
	\textbf{d,}
	Sign change of the exchange energy for $U<\hbar\omega$. Singlet fraction as a function of evolution time for the parameters in \textbf{c} with $U/h=6.5(1)\;\text{kHz}$ in the static case (black) and after a sudden switch on of the modulation with $K_0=0.88(1)$ (cyan) or $K_0=1.31(2)$ (orange) after a quarter exchange oscillation. Due to the sign reversal of $J_{\text{ex}}$, the rotation direction on the Bloch sphere is reversed. Solid lines are damped sine fits to the data. Error bars denote the standard deviation of 3 measurements.
	}
	\label{fig:4}
\end{figure} 

The results of the measurements are shown in Fig.\;4 in the off- and near-resonant driving regimes for a modulation frequency of $\omega/2\pi=8\;\text{kHz}$.
In the case of high frequency modulation with $t_x\ll U\ll\hbar\omega$, the tunneling is renormalized according to Eq.\;(2) and the exchange energy decreases as a function of the driving amplitude as $J_{\text{ex}}\approx 4 t_x^2 \mathcal{J}_0^2(K_0)/U$ (see Fig.\;4b). 
In contrast, in the near-resonant regime the system is to lowest order described by the tunneling process in Eq.\;(3) and we observe an increasing exchange energy as a function of the modulation strength for $\delta<0$ (see Fig.\;4c). 
At $K_0\approx 1.6$ it reaches a level about three times higher than in the static case. 
If $\delta>0$, $J_{\text{ex}}$ vanishes at a critical modulation amplitude of $K_0\approx 0.7$ and changes sign for stronger driving. 
To demonstrate that the exchange becomes negative for large $K_0$, we first perform a quarter oscillation in the static double well, followed by a sudden switch on of the modulation with $K_0>0.7$\;\cite{Trotzky2008}. 
Since the exchange in the driven double well is ferromagnetic, it inverts its rotation direction on the Bloch sphere, which leads to an oscillation phase shifted by $\pi$ compared to the static case (Fig.\;4d). 

The dependence of the exchange energy on the driving frequency and strength provide a microscopic explanation for the phenomena observed in the many-body system. 
In the off-resonant case, the magnetic exchange decreases with increasing modulation amplitude, which reduces the lattice anisotropy and therefore the correlations on the $x$-bonds (see Fig.\;2b). 
If the interaction energy $U$ comes close to, but is still lower than the driving frequency, resonant effects start to dominate and the magnetic exchange inverts its sign, leading to ferromagnetic correlations in the many-body system as observed in Fig.\;3b,c.  
For $U\gtrsim\hbar\omega$ the exchange energy increases with $K_0$, which can enhance anti-ferromagnetic correlations due to several reasons. 
First, the anisotropy is increased since the ratio $J_{\text{ex}}^x/J_{\text{ex}}^{y,z}$ becomes larger, which makes it more favorable to redistribute entropy onto the weak links in $y$- and $z$-directions\;\cite{Imriska2014,Greif2015}. 
Second, while the exchange is increased, the single-particle tunneling energy is renormalized as $t_{x,\text{single}}=t_x\mathcal{J}_0(K_0)$ in the effective Hamiltonian, see Eq.\;(3). 
Due to the isolated nature of the entire system, the reduction of $t_{x,\text{single}}$ leads to an entropy redistribution in the trap and lowers the absolute temperature, which globally enhances magnetic correlations. 
Last, when the ratio $J_{\text{ex}}/t_{x,\text{single}}$ increases, it becomes more favorable for two atoms to pair and form a singlet state in the low filled regions of the trap instead of delocalizing far apart\;\cite{Coulthard2016}. 
This process plays an important role in the context of high-temperature superconductivity, and the independent control of the exchange and tunneling energies opens up the possibility to investigate $d$-wave pairing in the $t$-$J$-model\;\cite{Dagotto1994}. 
Further theoretical studies will be necessary to determine the degree to which these three effects are responsible for the enhancement of anti-ferromagnetic correlations in the many-body system. 

Having shown that near-resonant driving can be used to increase or reverse the sign of magnetic correlations, the low energy scales in cold atom systems enable further investigations of the involved timescales and the possible existence of pre-thermalized states in future experiments\;\cite{Kuwahara2016}. 
Remarkably, the lifetime of correlations in the driven many-body system was found to be sufficiently
long to observe them even in the near-resonant driving regime. 
To investigate this further, the entropy increase could be systematically studied as a function of the involved energy scales and the connectivity of the underlying lattice geometry. 
Furthermore, by additionally imprinting complex phases on the density assisted tunneling energies, dynamical gauge fields and anyonic statistics could be engineered\;\cite{Bermudez2015}. 

\bibliographystyle{naturemag}

\vspace{0.5cm}

\textbf{Acknowledgements} 
We thank D. Abanin, D. Greif, D. Jaksch, M. Landini, Y. Murakami, N. Tsuji, P. Werner and W. Zwerger for insightful discussions. We acknowledge SNF (Project Number 200020\_169320 and NCCR-QSIT), Swiss State Secretary for Education, Research and Innovation Contract No.\;15.0019 (QUIC) and ERC advanced grant TransQ (Project Number 742579) for funding. 

\textbf{Author Contributions} 
All authors contributed extensively to the work presented in this manuscript. 

\textbf{Author Information} 
The authors declare no competing financial interests. Correspondence and requests for materials should be addressed to T.E. (esslinger@phys.ethz.ch). 

\clearpage

\makeatletter
\setcounter{section}{0}
\setcounter{subsection}{0}
\setcounter{figure}{0}
\setcounter{table}{0}
\setcounter{NAT@ctr}{0}

\renewcommand{\tablename}{Extended Data Table}
\renewcommand{\thetable}{ED\@arabic\c@table}

\renewcommand{\figurename}[1]{Extended Data Figure }
\renewcommand{\thefigure}{ED\@arabic\c@figure}

\section*{METHODS} 
\subsection{Optical lattice}
The tunable three-dimensional optical lattice is created by a combination of four orthogonal, retro-reflected laser beams of wavelength $\lambda=1064\,\mathrm{nm}$, as shown in Fig.\;1a.
While the $\overline{X}$ and $\widetilde{Y}$ beams are interfering and actively phase locked to $\varphi=0.00(3)\pi$, the  $X$ and $Z$ beams are non-interfering due to a frequency detuning. 
Our optical set-up is described by the following potential\;\cite{sTarruell2012}:  
\begin{eqnarray} V(x,y,z) & = & -V_{\overline{X}}\cos^2(kx+\theta/2)-V_{X} \cos^2(kx) \notag\\
&&-V_{\widetilde{Y}} \cos^2(k y) -V_{Z} \cos^2(k z) \\
&&-2\alpha \sqrt{V_{X}V_{Z}}\cos(k x)\cos(kz)\cos\varphi \notag
\label{Lattice}
\end{eqnarray}
with $k=2\pi/\lambda$ and $V_{\overline{X},X,\widetilde{Y},Z}$ as the lattice depths in units of the recoil energy $E_{\text{R}}=h^2/2m\lambda^2$ of each laser beam in the three different directions $x,y,z$  ($h$ is the Planck constant and $m$ the mass of the atoms). 
The lattice potential is adjusted to fix $\theta=\pi\times 1.000(2)$. 
We calibrate the visibility of the interference term $\alpha=0.92(1)$ with amplitude modulation of the lattice depth for different configurations of the optical potential using a $^{87}\mathrm{Rb}$ Bose-Einstein condensate. 
To calibrate the individual lattice depths $V_{\overline{X},X,\widetilde{Y},Z}$ we perform Raman-Nath diffraction on the Bose-Einstein condensate. 
For the calculation of tight-binding parameters, we include a systematic error of 3\% for all lattice depths. 
\subsection{Preparation of the degenerate Fermi gas in the optical lattice} 
The starting point of our experiment is a balanced mixture of the $F=9/2, m_F=-9/2$ and $F=9/2,m_F=-7/2$ hyperfine states of $^{40}\text{K}$, confined in an optical harmonic trap. 
We evaporatively cool the mixture to a quantum degenerate cloud with a repulsive s-wave scattering length of $115.6(8)\;a_0$ ($a_0$ denotes the Bohr radius). 
After the evaporation, we end up with about $3.0(2)\times 10^4$ (10\% systematic error) atoms at a temperature of $T/T_{\text{F}}=0.07(1)$ ($T_{\text{F}}$ denotes the Fermi temperature, see extended data Tab.\;ED1 for details). 
Afterwards, we either keep a mixture of the $F=9/2, m_F=-9/2$ and $F=9/2,m_F=-7/2$ hyperfine states to access attractive or weak repulsive interactions with scattering lengths $-3000\;a_0<a<150\;a_0$ (measurements in Figs.\;2 and 4b and for the initial preparation of isolated double wells in Fig.\;4). 
Alternatively, we transfer the $F=9/2, m_F=-7/2$ state to the $F=9/2, m_F=-5/2$ state with a radio-frequency sweep to access large repulsive scattering lengths above $200\;a_0$ (measurements in Figs.\;3 and 4c,d). 
For this mixture, we obtain temperatures of $T/T_{\text{F}}=0.12(2)$ in the harmonic trap. 
The interactions can be tuned via two magnetic Feshbach resonances located at a field of $202.1\;\text{G}$ (for $m_F=-9/2$ and $m_F=-7/2$) and $224.2\;\text{G}$ (for $m_F=-9/2$ and $m_F=-5/2$), respectively. 
From this point, two distinct schemes are used to either prepare atoms in a three-dimensional hexagonal lattice (Figs. 2, 3) or in isolated double wells (Fig.\;4). 
To load a many-body state into the hexagonal lattice, we first ramp up the power of all lattice beams in $50\;\text{ms}$ to an intermediate value. In this configuration, the tunneling energies are close to the final configuration with $(t_x,t_y,t_z)/h=(550(30),143(8),156(9))\;\text{Hz}$ but the horizontal link across the hexagonal unit cell has still a finite value of $70(3)\;\text{Hz}$. In addition, the mean trap frequency is only $\bar{\omega}_{\text{trap}}=68(2)\;\text{Hz}$. 
In a second step, we ramp up the power in all beams in $20\;\text{ms}$ to the final configuration (see extended data Tab.\;ED1 for the detailed values). 
To load isolated double wells, we first tune the interactions to a large attractive value of $-3000(600)\,a_0$, see \cite{Desbuquois2017} for more details. 
In short, the atoms are first loaded into the lowest band of a checkerboard configuration with $V_{\overline{X},X,\widetilde{Y},Z}=[0,3,7,3]\;E_{\text{R}}$ using an S-shaped lattice ramp of $200\;\text{ms}$. 
Due to the large attractive interactions during the loading process, 68(3)\% of the atoms form double occupancies. 
In a second step we then tune the scattering length to $115.6(8)\;a_0$ and split each lattice site by linearly increasing $V_{\overline{X}}$ and decreasing $V_X$ to a $V_{\overline{X},X,\widetilde{Y},Z}=[30,0,30,30]\;E_{\text{R}}$ cubic configuration within $10\;\text{ms}$. 
During the splitting process, the double occupancies in the checkerboard lattice are transformed into singlet states $\left|\text{s}\right\rangle=(\left|\uparrow,\downarrow\right\rangle-\left|\downarrow,\uparrow\right\rangle)/\sqrt{2}$ in the cubic lattice. 
\subsection{Detection methods}
The detection scheme of double occupancies and nearest neighbor spin-spin correlations follows closely the procedure used in earlier work (see\;\cite{sGreif2013,Greif2015}). 
To characterize the atomic state, we first freeze the evolution by quenching the lattice to $V_{\overline{X},X,\widetilde{Y},Z}=[30,0,30,30]\;E_{\text{R}}$ within $100\;\mu\text{s}$. 
In order to detect double occupancies, we ramp the magnetic field close to the magnetic Feshbach resonance of the $m_F=-9/2$ and $m_F=-7/2$ mixture. 
We then selectively transfer one of the atoms sitting on doubly occupied sites from the $m_F=-7/2$ state, to the $m_F=-5/2$ state or vice versa via a radio frequency sweep by making use of the interaction shift. 
The number of atoms in the different Zeeman sublevels can then be determined by applying a Stern-Gerlach pulse during the time-of-flight imaging. 
For the measurement of spin-spin correlations, we apply a magnetic field gradient after the lattice freeze. 
This leads to coherent oscillations between the magnetic singlet state $\left|\text{s}\right\rangle=(\left|\uparrow,\downarrow\right\rangle-\left|\downarrow,\uparrow\right\rangle)/\sqrt{2}$ and triplet state $\left|\text{t}\right\rangle=(\left|\uparrow,\downarrow\right\rangle+\left|\downarrow,\uparrow\right\rangle)/\sqrt{2}$ on neighboring sites in the $x$-direction. 
The singlet fraction $p_{\text{s}}$ can be determined by merging adjacent lattice sites by going to a $V_{\overline{X},X,\widetilde{Y},Z}=[0,30,30,30]\;E_{\text{R}}$ checkerboard configuration within $10\;\text{ms}$. 
This procedure transforms the singlet into a double occupancy in the single well, which can again be measured as outlined above. 
The triplet fraction $p_{\text{t}}$ is obtained by applying a $\pi$-pulse with the magnetic field gradient and subsequently measuring the singlet fraction. 
The spin-spin correlator is then obtained as $\mathcal{C}=-\langle \hat{S}_{i}^x \hat{S}_{i+1}^x\rangle-\langle \hat{S}_{i}^y \hat{S}_{i+1}^y\rangle=(p_{\text{s}}-p_{\text{t}})/2$.
We average all observables over one period $T=2\pi/\omega$ of the drive to be insensitive to the micromotion.
For that, we vary slightly the total duration of the modulation between different measurements by multiples of $T/4$, in order to sample different phases of the modulation cycle. 
For the measurement of double occupancies in the hexagonal lattice (Figs. 2a and 3a), we sample four different times during the modulation cycle, while for the magnetic correlations (Figs. 2b, 3b, 3c and extended data Figs.\;ED1, ED2) we measure for five different times and take each data point two or three times (for exact number of measurements see captions). 
For the measurements performed in the isolated double wells (Fig.\;4) the observables were not averaged over one driving period, since we have experimentally verified that no fast dynamics could be observed in this configuration. 
This can be explained by $\hbar\omega$ being much larger than $t$. 
\subsection{Periodic driving}
The periodic driving is implemented as in previous work \cite{Desbuquois2017}. In brief, a piezo-electric actuator allows for a controlled phase shift of the reflected $X$ and $\overline{X}$ lattice beam with respect to the incoming beams.
To access the driven regime, we modulate the lattice position by a sinusoidal movement of the mirror position for the retro-reflecting lattice beam at frequency $\omega/2\pi$.
We choose the modulation to be along the direction of the horizontal bonds such that $V(x,y,z,\tau)\equiv V(x-A\cos(\omega\tau),y,z)$.
In a first step we linearly ramp up a sinusoidal modulation and then maintain a fixed displacement amplitude $A$.
During the modulation we ensure the correct phase relation $\varphi=0.0(1)\pi$ between the two interfering X and Z lattice beams by modulating the phase of the respective incoming beams at the same frequency using acousto-optical modulators.
In addition, this phase modulation is used to calibrate the phase and amplitude of the mirror displacement. 
In our setup the piezo modulation also leads to a residual periodic reduction of the interference amplitude of the lattice by at most $2\,\%$. For the lattice configurations used in our experiments, this shifts the mean tunneling energy $t_x$ down by about $2.5\,\%$ and introduces a modulation of the tunneling energy at twice the driving frequency $2\omega/(2\pi)$ with an amplitude of $\delta t=0.025\:t_x$. The effect of the modulation is negligible, since its amplitude has to be compared to the driving frequency. The effective driving strength is given by $\delta t/(2\hbar\omega)$, which is always smaller than $3\times 10^{-3}$ in our case. 
In addition, we have verified that our experimental findings are not affected by the launching phase of the drive.
The amplitude of the lattice displacement $A$ is directly related to the normalized drive amplitude by $K_0=m\,A\,\omega\,d_x/\hbar$, where $d_x$ is the distance between the two sites along the $x$-direction.
For our lattice potential, $d_x \neq \lambda/2$, and must be calculated for each individual configuration.
To this end, we determine the Wannier functions located on the left and right sides of the considered bond, which are derived as the eigenstates of the band-projected position operator.
The distance $d_x$ is then evaluated as the difference between the eigenvalues of two neighboring Wannier states, and is given in extended data Tab.\;ED1 for all lattice configurations. 
In addition, since the lattice geometry in the $x$-$z$-plane is not an ideal brick configuration, the bonds connecting two sites in the $z$-direction are also slightly affected by the drive. 
The effective driving strength can be determined by the projected bond length on the modulation direction, which for our case is the $x$-displacement $d_x^{\text{vert}}=\lambda/2-d_x$ between neighboring sites in the vertical $z$-direction. 
The modulation amplitude is then given by $K_0^{\text{vert}}=d_x^{\text{vert}}/d_x K_0$. 
The values for $d_x^{\text{vert}}$ are given in extended data Tab.\;ED1 for our lattice configurations.
\subsection{Calibration of the on-site interactions}
The extension of the Wannier function can be similar to the scattering length for strong interactions in the optical potentials realized in our measurements. Thus, the actual on-site interaction strength $U$ may be altered compared to the value calculated with the non-interacting Wannier functions as observed in previous experiments\;\cite{sUehlinger2013,Desbuquois2017}.
We therefore determine $U$ experimentally by driving the lattice at a frequency $\omega/2\pi$, and measure the amount of double occupancies as a function of $U$. 
Double occupancies are maximally created either for $\hbar\omega = U$ in a connected lattice (Figs.\;2 and 3), or for $\hbar\omega = (\sqrt{U^2+16\,t^2}+U)/2$ in the isolated double wells (Fig.\;4). 
In the hexagonal lattice, the resonance position is within agreement of the numerical value for $U$ determined from the Wannier function, as shown in Fig.\;3a. 
However, a significant difference is observed in the isolated double wells. 
To account for this effect, we parametrize $U$ by $U(a) = \alpha\,a\,(1-a/a_c)$, where $\alpha$ is given by the non-interacting Wannier functions and $a_c$ is a higher order correction and depends on the lattice depth. 
For the isolated double wells, we find $a_c = 4800\, (300)\,a_0$, leading to a reduction in $U$ of about $10\;\%$ with respect to the calculated value for the data sets shown in Fig.\;4c. 
Accordingly, this correction is incorporated to all interaction strengths given for the isolated double wells. 
\subsection{Validity of tight-binding approximation and higher band effects}
When deriving the tight-binding Hamiltonian of the driven Fermi-Hubbard model Eq.\;(1), it is assumed that the Wannier functions are not modified by the modulation. However, for large driving amplitudes a significant tilt is applied to the lattice in the co-moving frame, which introduces an energy bias $\hbar\omega K_0$ between neighboring sites (see also Fig.\;1). As a result, the Wannier functions will be modified due to the admixture of higher-band Wannier functions of the untilted lattice. This will in turn lead to different tight-binding parameters $t_x$ and $U$ at any given time within the modulation cycle. In order to estimate the corrections resulting from the change of Wannier functions, we consider a cut through the tilted lattice potential in the $x$-direction of the modulation. This potential can be very well approximated around the horizontal bonds by a lattice with a relative phase $\theta\neq\pi$ between the lattice beams $\overline{X}$ and $X$ (see Eq.\;(4)). The approximation in this step is to assume that all lattice sites in a given sublattice ($\mathcal{A}$ or $\mathcal{B}$) are at equal energy. This is well justified for our lattice geometry, since the tunneling energy across the hexagon is zero and thus the Wannier functions e.g. on the $\mathcal{A}$ sublattice are not influenced by the $\mathcal{B}$ sites to their left. Since the discrete spatial periodicity is restored in the lattice potential with $\theta\neq\pi$, we can compute the Wannier functions for any given energy bias and calculate the corresponding tight-binding parameters. The modulated lattice potential can then be described by a tight-binding Hamiltonian as in Eq.\;(1), where in addition to the oscillating force $f(\tau)$ the Hubbard parameters $t_x(\tau)$ and $U_{\mathcal{A},\mathcal{B}}(\tau)$ become time- and sublattice-dependent. We decompose the parameters into their Fourier components, which take the form 
\begin{eqnarray}
t_x(\tau) &=& t_x(K_0=0)+\delta t_0(K_0,\omega)+\delta t_2(K_0,\omega)\cos(2\omega\tau) \notag\\ 
&& +... \notag\\
U_{\mathcal{A}}(\tau) &=& U(K_0=0)+\delta U_0(K_0,\omega)+\delta U_1(K_0,\omega)\cos(\omega\tau) \notag\\
&&+\delta U_2(K_0,\omega)\cos(2\omega\tau)+... \notag\\
U_{\mathcal{B}}(\tau) &=& U_{\mathcal{A}}(\tau+\pi/\omega). \notag
\end{eqnarray}
The expansion of $t_x(\tau)$ features only even harmonics of $\omega$ since $t_x(\tau)=t_x(\tau+\pi/\omega)$. The main effect of the modulation is a shift of the static tunneling energy by $\delta t_0(K_0,\omega)$, which is given in extended data Tab.\;ED1 for the maximum driving amplitude and frequency in each lattice configuration. Note that even though the relative change of the tunneling energy is on the order of 10-20\% for large values of $K_0$, the absolute change is much smaller since the hopping amplitude is renormalized by the Bessel functions $\mathcal{J}_{0}(K_0)$ or $\mathcal{J}_{1}(K_0)$, depending on the frequency regime. On the other hand, we find that the shift of the mean value of $U$ is much smaller and even for the strongest driving we have $\delta U_0(K_0,\omega)/U<6\times 10^{-3}$. The second effect is a modulation of $t_x$ and $U$ which is negligible, since it has to be compared to the driving frequency. The dimensionless modulation strength for the lowest Fourier components will be given by $K_0^t=\delta t_2(K_0,\omega)/(2\hbar\omega)$ and $K_0^U=\delta U_1(K_0,\omega)/(\hbar\omega)$. Even for the maximum values of $K_0$ and $\omega$, we find $K_0^t<6\times 10^{-3}$ and $K_0^U<0.02$ for all our lattice geometries. We have also performed a numerical simulation of the two-site Hubbard model including all of the above modifications, in which we use a Trotter decomposition to evaluate the quasi-energy spectrum (see also Methods section on the theoretical treatment of the driven double well and extended data Fig.\;ED4). We have found that even for the largest driving amplitudes used in the measurement of the exchange energy (see Fig.\;4), $J_{\text{ex}}$ is modified by at most $10\;\text{Hz}$ in the off-resonant driving regime (compare to extended data Fig.\;ED4b) and $60\;\text{Hz}$ in the near-resonant case (extended data Fig.\;ED4d), which is mainly caused by the shift of the mean value of $t_x$. This change is still smaller or comparable to the uncertainty on the exchange energy resulting from an imprecise calibration of the Hubbard parameters in the lattice, which is on the order of $70\;\text{Hz}$. 

\subsection{Measurement of magnetic exchange}
The exchange energy is measured in a Ramsey-type protocol in isolated double wells. 
After preparing singlet states on adjacent sites in a deep cubic lattice with $V_{\overline{X},X,\widetilde{Y},Z}=[30,0,30,30]\;E_{\text{R}}$ as outlined above, we perform a $\pi/2$-pulse with a magnetic field gradient to generate a coherent superposition between the singlet and triplet state. 
After this, we first ramp the magnetic field, the interfering lattice $V_X$ and the driving amplitude $K_0$ to the desired value within $2\;\text{ms}$. 
In the next step, we trigger an exchange oscillation by suddenly lowering the barrier in the double well by decreasing $V_{\overline{X}}$ to the desired value within $100\;\mu\text{s}$. 
After a variable evolution time $\tau_{\text{evol}}$ in the driven system, we freeze the dynamics again by increasing $V_{\overline{X}}$ to $30\;E_{\text{R}}$ within $100\;\mu\text{s}$, revert the the ramps of the magnetic field, the interfering lattice $V_X$ and the driving amplitude $K_0$ and perform a second $\pi/2$-pulse with a magnetic field gradient. 
Finally, we measure the fraction of singlets on adjacent sites, which is given by $p_{\text{s}}(\tau_{\text{evol}})=[1-\cos(J_{\text{ex}}\tau_{\text{evol}}/\hbar)]/2$ after the evolution. 
In the experiment, we vary the evolution time $\tau_{\text{evol}}$ and measure the singlet fraction for each modulation amplitude $K_0$ for not less than 9 different values of $\tau_{\text{evol}}$, with at least 27 individual measurements in total. We fit the data with a function $p_{\text{s}}(\tau_{\text{evol}})=\alpha [1-\cos(J_{\text{ex}}\tau_{\text{evol}}/\hbar)]\exp[-\beta\tau]+\gamma$ and extract the exchange from the fitted frequency. In order to estimate the error, we use a resampling method which assumes a normal distribution of measurement results at each evolution time. The standard deviation of the distribution is determined by the measured standard deviation or, if we measured the singlet fraction at this $\tau_{\text{evol}}$ only once, by the residual from the fitted curve. Afterwards, we randomly sample a value for the singlet fraction at each evolution time and refit the resulting data set. At the same time, the initialization values of the fit parameters $J_{\text{ex}}$ and $\beta$ are varied by $\pm 10\%$. This procedure is repeated 1000 times and the mean $\pm$ standard deviation of the resulting distribution of frequencies determine the asymmetric error bars for the fitted exchange frequency, as shown in Fig.\;4.
In order to demonstrate the sign change of the magnetic exchange for $U\lesssim\hbar\omega$ (see Fig.\;4d), we first let the system evolve for a time $\tau_0$ with a non-driven exchange $J_{\text{ex}}^{(0)}$, until a quarter exchange oscillation has been performed, i.e. $J_{\text{ex}}^{(0)}\tau_0=\pi/2$. 
After that, we suddenly switch on the sinusoidal modulation at the desired value of $K_0$, which projects the system on to a Hamiltonian with a negative $J_{\text{ex}}$. 
Therefore, the system changes its sense of rotation on the Bloch sphere (see Fig.\;4a) and the singlet fraction after a variable total evolution time $\tau_{\text{evol}} > \tau_0$ is given by $p_{\text{s}}(\tau_{\text{evol}})=\{1+\text{sgn}(J_{\text{ex}})\sin[\left|J_{\text{ex}}\right|(\tau_{\text{evol}}-\tau_0)/\hbar]\}/2$.
\subsection{Theoretical treatment of the driven double well}
We perform both analytic and numerical studies on the driven double well as described in earlier work\;\cite{Desbuquois2017}. In this context, we use Floquet's theorem to derive an effective static Hamiltonian in a high-frequency expansion. In the following, we will include terms up to order $1/\omega$, as given in Appendix\;A in \cite{Desbuquois2017}. 
In the off-resonant case, the term proportional to $1/\omega$ vanishes, such that the effect of the modulation is a pure renormalization of the tunneling by a 0-th order Bessel function $t\rightarrow t\mathcal{J}_0(K_0)$. Therefore, the exchange energy defined as the energy difference between the triplet and singlet state becomes
\begin{equation}
J_{\text{ex, off-res}}=\frac{1}{2}\left[-U+\sqrt{16 t^2\mathcal{J}_0^2(K_0)+U^2}\right] \notag
\end{equation}
In the Heisenberg limit of large interactions ($t\ll U\ll \hbar\omega$) we find
\begin{equation}
J_{\text{ex,off-res}}\stackrel{U\gg t}{\longrightarrow} 4\frac{t^2}{U}\mathcal{J}_0^2(K_0)
\label{HeisenbergLimit}
\end{equation}
In the case of near-resonant driving ($t\ll U\approx \hbar\omega$), we can express the Hamiltonian in terms of $t$, $U$ and the detuning $\delta=\hbar\omega-U$ and we consider terms up to orders $\mathcal{O}(t^2/U,t\delta/U,\delta^2/U)$. In this regime, the single particle tunneling $t_0=t\mathcal{J}_0(K_0)$ is renormalized as for the off-resonant case. On the other hand, the density assisted tunneling changing the number of double occupancies is given by $t_1=t\mathcal{J}_1(K_0)$. The exchange is given by
\begin{equation}
J_{\text{ex, res}}=\frac{1}{2}\left[\delta+4\frac{t_0^2}{U}\mp\sqrt{16 t_1^2+\left(\delta-4\frac{t_0^2+t_1^2}{U}\right)^2}\right] \notag
\end{equation}
for $\delta\gtrless 0$, which reproduces the Heisenberg limit Eq.\;(5) for the case of no driving $K_0=0$. For large detunings ($t\ll \delta\ll U,\hbar\omega$), the exchange takes the form
\begin{equation}
J_{\text{ex,res}}\stackrel{\delta\gg t}{\longrightarrow}-4\frac{t_1^2}{\delta}+2\frac{2t_0^2+t_1^2}{U} \notag
\end{equation}
The leading term of this expansion is proportional to $\mathcal{J}^2_1(K_0)$ and changes sign with the detuning $\delta$. This explains the switch to a ferromagnetic exchange for $U<\hbar\omega$ beyond a certain driving strength.
In addition to the analytic derivation of the effective Hamiltonian, we also performed a numerical simulation of the two-site Hubbard model. We use a Trotter decomposition to evaluate the evolution operator over one period from which we extract the spectrum (for details see \cite{Desbuquois2017}). A comparison of the numerical and analytic results for the experimental parameters is shown in extended data Fig.\;ED4.
For all of the derivations above, we assumed that the static double well can simply be described by the tunneling $t$ and the onsite interaction $U$. However, if the Wannier functions on the two sites have a significant overlap, the description needs to be extended to a two-band Hubbard model. In this case, higher order corrections like density assisted tunneling $\delta t$ as well as nearest neighbor interactions, direct exchange and correlated pair tunneling $V$ (the last three are all equal for the two-band Fermi-Hubbard model) become significant (see Appendix\;A.1 in \cite{Desbuquois2017}). For the experimental parameters in the off-resonant case (see Fig.\;4b), their values are $V/h=2.4(7)\;\text{Hz}$, $\delta t/h=22(3)\;\text{Hz}$ in the static lattice. In the near-resonant driving regime (Fig.\;4c), interactions are stronger and the corrections increase to $V/h=26(8)\;\text{Hz}$, $\delta t/h=120(10)\;\text{Hz}$ for $U/h=6.5(1)\;\text{kHz}$ and $V/h=40(10)\;\text{Hz}$, $\delta t/h=170(20)\;\text{Hz}$ for $U/h=9.1(1)\;\text{kHz}$. 
To lowest order, the density assisted tunneling will increase the effective tunneling to be $t+\delta t$, and $V$ decreases the exchange interaction by $2V$, both in the static and driven cases.
\subsection{Data availability}
All data files are available from the corresponding author upon request.

\clearpage

\section*{ }

\begin{figure}
	\includegraphics[width=1\columnwidth]{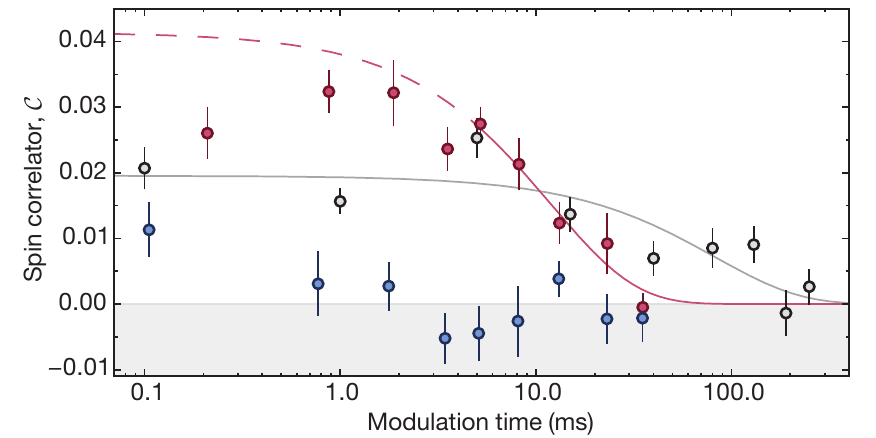}
	\caption{
		\textbf{Time dependence of magnetic correlations for near-resonant driving.} 
	Nearest neighbor spin-correlations $\cal{C}$ for the same lattice configuration as in Fig.\;3 as a function of the modulation time after the ramp up of the drive. The data allows to compare the formation and decay of magnetic correlations for two specific sets of interactions and modulation frequencies with the level of correlations in the static case (black). For a driving strength of $K_0=1.30(3)$ and $U/h =3.8(1)\;\text{kHz}$, $\omega/2\pi=3\;\text{kHz}$ (red), anti-ferromagnetic correlations increase with time and reach a level higher than the static case (black, $U/h=3.8(1)\;\text{kHz}$). If the interaction is smaller than the driving frequency (blue, $U/h=4.4(1)\;\text{kHz}$, $\omega/2\pi=6\;\text{kHz}$), the correlations switch sign and become ferromagnetic after a few milliseconds. For long times, the correlations in each configuration decrease due to heating in the lattice. Solid lines show an exponential fit of the full data in the static case (gray) and to modulation times longer than $4\;\text{ms}$ in the driven lattice for $U>\hbar\omega$ (red). The extracted lifetimes decrease from $82(34)\;\text{ms}$ without drive to $12(4)\;\text{ms}$ at $K_0=1.30(3)$. All measurements are averaged over one modulation cycle. Data points and error bars denote the mean and standard error of 13 individual measurements at different times withing one driving period (see Methods). 
	}
	\label{fig:S1}
\end{figure}

\begin{figure}
	\includegraphics[width=1\columnwidth]{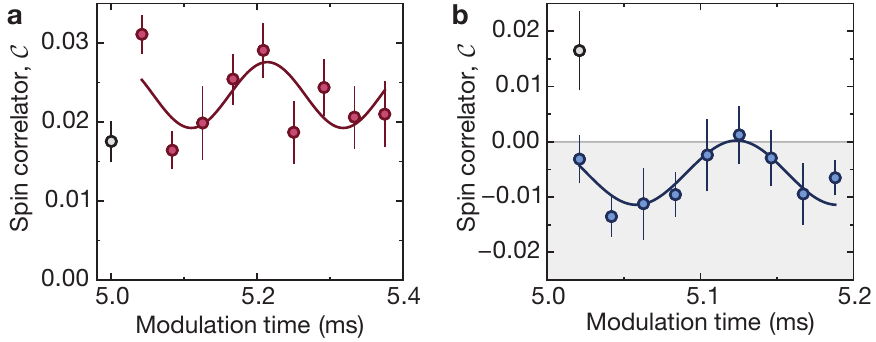}
	\caption{
	\textbf{Micromotion for near-resonant driving.} 
	Nearest neighbor spin-correlations $\cal{C}$ for the lattice configuration in Fig.\;3 and $K_0=1.30(3)$ as a function of modulation time after the ramp up of the drive, sampled withing one oscillation period. We observe a significant micromotion both for the case of enhanced anti-ferromagnetic correlations in \textbf{a} ($U/h=3.8(1)\;\text{kHz}$ and $\omega/2\pi=3\;\text{kHz}$) and for ferromagnetic ordering in \textbf{b} ($U/h=4.4(1)\;\text{kHz}$ and $\omega/2\pi=6\;\text{kHz}$). For a different set of parameters in the measurement of the micromotion it should be also possible to switch between anti-ferromagnetic and ferromagnetic correlations within one driving cycle. The open symbols represent a reference measurement in the static case with all other parameters being equal. Solid lines are sinusoidal fits to the data which results in a fitted frequency of $4.8^{+1.9}_{-0.4}\;\text{kHz}$ (\textbf{a}) or $7.6^{+3.9}_{-1.7}\;\text{kHz}$ (\textbf{b}), respectively. Error bars denote the standard error of 10 independent measurements.
	}
	\label{fig:S2}
\end{figure} 

\begin{figure}
	\includegraphics[width=1\columnwidth]{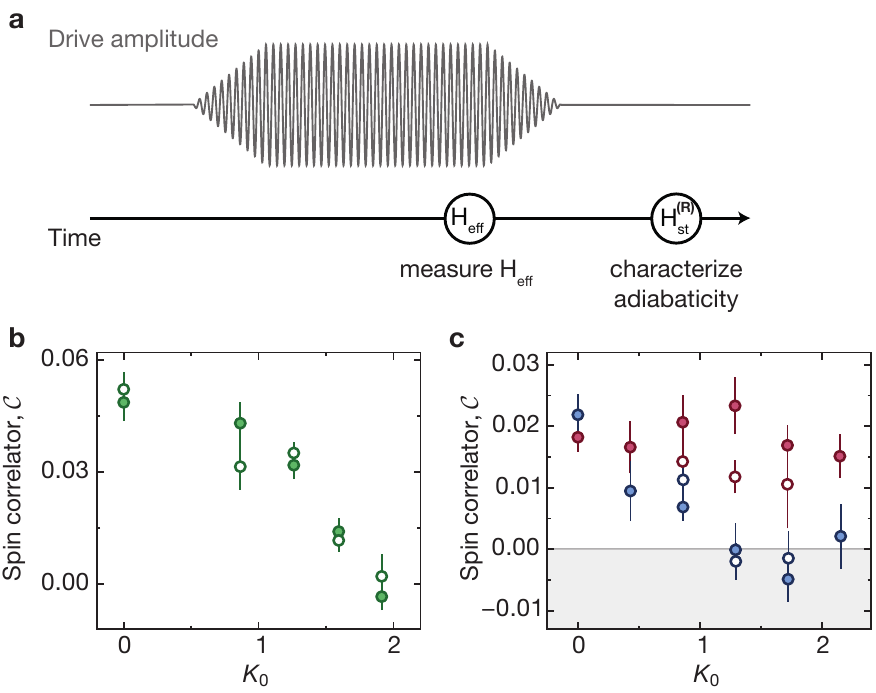}
	\caption{	
	\textbf{Adiabaticity of the modulation ramp in the many-body system.}
	\textbf{a,} Starting from the static lattice, the modulation amplitude is ramped up and subsequently kept at a fixed value to allow for a $5\;\text{ms}$ equilibration time. The ramp up time depends on the chosen configuration and is $3.333\;\text{ms}$ ($2\;\text{ms}$) for a modulation frequency of $\omega/2\pi=3\;\text{kHz}$ ($6\;\text{kHz}$). We start the detection of nearest neighbor spin-correlations $\cal{C}$ by quenching the tunneling to zero as we ramp up the lattice depth in all directions within $100\;\mu\text{s}$. To estimate the adiabaticity of the final state, we perform a second type of measurement in which we revert the driving ramp followed by an additional waittime of $5\;\text{ms}$ before the detection. If the ramp scheme of the modulation is fully adiabiatic, we expect a reversal of the correlations to their static value. 
\textbf{b,}  The nearest neighbor spin-correlations $\cal{C}$ are plotted versus the modulation strength in the off-resonant driving regime ($U/h=0.93(2)\;\text{kHz}$ and $\omega/2\pi=6\;\text{kHz}$). The filled green circles are measured in the modulated system (same data as in Figure 2b) and the open green circles after ramping off the modulation. It can be observed that the correlations do not reach the level of the static case at $K_0=0$ anymore after reverting the ramp. We attribute this to some extend to a reduced lifetime of correlations, which is found to be $14(5)\;\text{ms}$ at $K_0=1.26(4)$ compared to $92(16)\;\text{ms}$ in the static case. 
\textbf{c,} Spin-correlator for different driving strengths $K_0$ in the near-resonant regime for $U<\hbar \omega$ (blue, $U=4.4(1)\;\text{kHz}$ and $\omega/2\pi=6\;\text{kHz}$) and in the regime of enhanced anti-ferromagnetic correlations (red, $U/h=3.8(1)\;\text{kHz}$ and $\omega/2\pi=3\;\text{kHz}$). Full data points represent the effective states in the modulated system (same data as in Figure 3c) while open data points are measured after ramping off the modulation. Again, correlations do not reach the static value after reverting the driving ramp due to the finite lifetime (see also extended data Fig.\;ED1). Data points and error bars denote the mean and standard error of 10 individual measurements at different times withing one driving period (see Methods). 
	}
	\label{fig:S3}
\end{figure}

\begin{figure}
	\includegraphics[width=1\columnwidth]{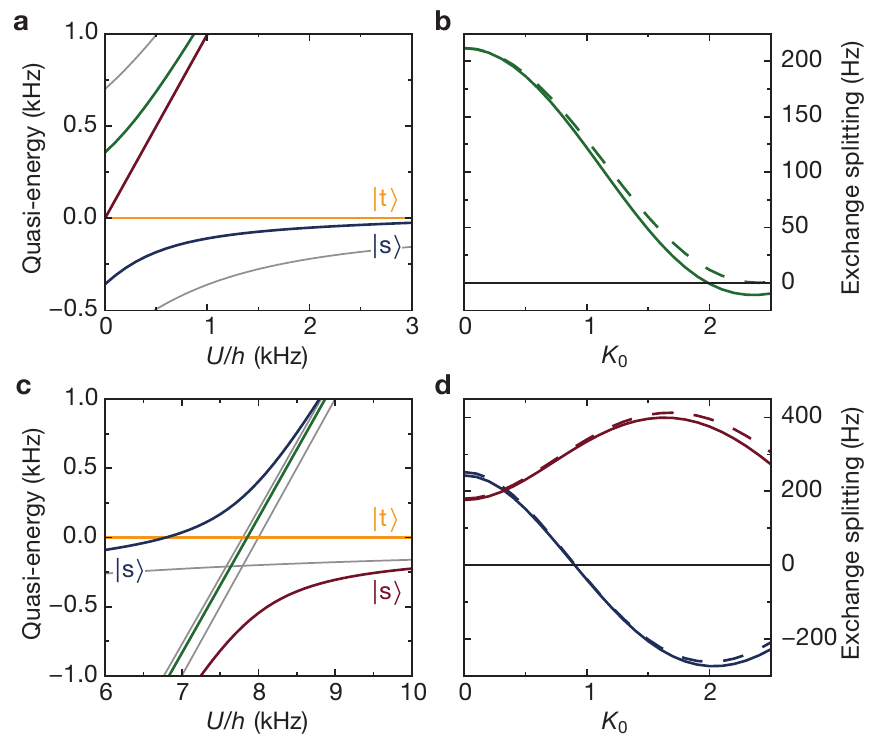}
	\caption{	
	\textbf{Analytical and numerical treatment of a driven double well.}
	\textbf{a,} Quasi-energy spectrum for two particles in a double well as a function of the onsite interaction $U$ for off-resonant driving ($t/h=350\:\text{Hz}$, $K_0=1.5$ and $\omega/2\pi=8\;\text{kHz}$). The gray lines show the energy spectrum without modulation. For $U\gg t$, the ground state is the spin singlet $\left|\text{s}\right\rangle$ and the first excited state the triplet $\left|\text{t}\right\rangle$. To lowest order, the driving renormalizes the tunneling by a zeroth order Bessel function $t_x\rightarrow t_x^{\text{eff}}(K_0)=t_x \mathcal{J}_0(K_0)\approx 0.51\;t_x$. 
	\textbf{b,} Calculated exchange energy $J_{ex,\mathrm{off-res}}$ (see Methods), defined as the energy difference between the spin singlet and triplet states (see \textbf{a}), as a function of the shaking strength $K_0$ for an off-resonant modulation ($t/h=350\;\text{Hz}$, $U/h=2.1\;\text{kHz}$ and $\omega/2\pi=8\;\text{kHz}$, compare to Fig.\;4b). The dashed line is the analytical result derived from a high-frequency expansion of the effective Hamiltonian, while the solid line is the result of a numerical calculation. The exchange energy is reduced to small values as the tunneling is renormalized by the zero-order Bessel function $\mathcal{J}_0(K_0)$. For large modulation amplitudes, deviations from the result obtained from an expansion up to order $1/\omega$ can be observed. Here, the exchange already becomes weakly ferromagnetic due to the finite value of the interaction. 
		\textbf{c,} Floquet spectrum of the double well system as a function of the interactions $U$ for near-resonant driving ($t/h=640\;\text{Hz}$, $K_0=0.8$ and $\omega/2\pi=8\;\text{kHz}$). The gray lines show the energy spectrum without periodic modulation. The drive couples the singlet state to a state containing double occupancy, which leads to an avoided crossing at $U\approx\hbar\omega$. As a result, a gap opens which is to lowest order given by $4\mathcal{J}_1(K_0)$. 
	\textbf{d,} Dependence of the exchange energy $J_{\text{ex,res}}$ on the modulation amplitude in the near-resonant regime for two different detunings with $t/h=640\;\text{Hz}$ and $\omega/2\pi=8\;\text{kHz}$ (blue data: $U/h=6.5\;\text{kHz}$; red data: $U/h=9.1\;\text{kHz}$, compare to Fig.\;4c). The dashed line is the analytical result (see Methods) derived from a high-frequency expansion of the effective Hamiltonian, while the solid line is the result of a numerical calculation. For $U>\hbar \omega$, the exchange energy is significantly increased while it changes sign to a ferromagnetic behavior for $U<\hbar \omega$. In both driving regimes, the analytical result is in very good agreement with the exact numerics. Our measurements of the exchange energy in Fig.\;4 agree well on a qualitative level with the theoretical expectation. 
	}
	\label{fig:S4}
\end{figure}

\clearpage

\begin{table}[h!]
\centering
\begin{tabular}{|c||c|c|c|c|c|}
\hline
Main text figure & 2 & 3 & 4b & 4c, 4d \\
\hline
Atom number ($10^3$) & 28(2) & 32(2) & \multicolumn{2}{c|}{186(6)} \\[2pt]
Initial $T/T_{\text{F}}$ & 0.07(1) & 0.12(2) & \multicolumn{2}{c|}{0.06(1)} \\[2pt]
$\bar{\omega}_{\text{trap}}/2\pi$ (Hz) & 84(2) & 84(2) & \multicolumn{2}{c|}{119(2)}\\[2pt]
$t_{x}/h$ (Hz) & 810(150) & 570(110) & 350(50) & 640(90)\\[2pt]
$t_{y}/h$ (Hz) & 125(8) & 125(8) & \multicolumn{2}{c|}{$<1$}\\[2pt]
$t_{z}/h$ (Hz) & 78(8) & 85(8) & \multicolumn{2}{c|}{$<2$} \\[2pt]
$d_x/(\lambda/2)$ & 0.71(2) & 0.74(2) & 0.79(1) & 0.73(1)\\[2pt]
$d_x^{\text{vert}}/(\lambda/2)$ & 0.29(2) & 0.26(2) & 0.21(1) & 0.27(1) \\[2pt]
$\delta t_0(K_0^{\text{max}},\omega^{\text{max}})/t_x$ & 0.085(1) & 0.102(1) & 0.236(9) & 0.106(2) \\[2pt]
\hline
\end{tabular}
\caption{
\textbf{Summary of experimental parameters for the measurements in Fig.\;2-4 of the main text.}
Values given for Fig.\;2 correspond to the initial static configuration with $K_0=0$. The initial temperature is measured before loading the atoms into the lattice. $d_x$ is the length of the horizontal bonds, while $d_x^{\text{vert}}$ is the horizontal distance between two sites forming the vertical bonds in $z$-direction, resulting from a non-rectangular lattice unit cell. The effective modulation amplitude is given by the projection of each bond on the $x$-direction. $\delta t_0$ describes the change of the mean value of $t_x$ in the driven lattice due to a time-dependent modification of the Wannier functions. The values given here are an upper bound corresponding to the maximum modulation amplitude $K_0^{\text{max}}$ and frequency $\omega^{\text{max}}$ used in each lattice configuration (see Methods for further details).}
\label{table:atnum}
\end{table}

\end{document}